\documentclass[prl,preprint,showpacs,showkeys,preprintnumbers,amsmath,amssymb]{revtex4-1}

\usepackage{graphicx}
\usepackage{dcolumn}
\usepackage{bm}

\begin{document}

\title{Orthogonality Catastrophe in Quantum Sticking}

\author{Dennis P. Clougherty}
\email{dpc@physics.uvm.edu}
\author{Yanting Zhang}

\affiliation{
Department of Physics\\
University of Vermont\\
Burlington, VT 05405-0125}

\date{\today}

\begin{abstract}
The probability that a particle will stick to a surface is fundamental to a variety of processes in surface science, including catalysis, epitaxial growth, and corrosion.  At ultralow energies, how particles scatter or stick to a surface affects the performance of atomic clocks, matter-wave interferometers, atom chips  and other quantum information processing devices. 
In this energy regime, the sticking probability is influenced by a distinctly quantum mechanical effect: quantum reflection, a result of matter wave coherence, suppresses the probability of finding the particle near the surface and reduces the sticking probability.   We find that another quantum effect can occur, further shaping the sticking probability: the orthogonality catastrophe, a result of the change in the quantum ground state of the surface in the presence of a particle, can dramatically alter the probability for quantum sticking and create a superreflective surface at low energies.
\end{abstract}

\maketitle

Recent experimental advances in the manipulation of ultracold atoms and molecules have provided new tools for investigating fundamental concepts in quantum many-body physics.  Experiments with ultracold particles \cite{bloch,zwierlein,greiner} now go well beyond simple studies of matter wave interference and probe the nature of superfluids, superconductors and Mott insulators.  Ultracold particles present a convenient way to probe the physics near quantum phase transitions, as coupling strengths are experimentally adjustable in many cases.

We demonstrate that another major concept from many-body physics--the orthogonality catastrophe--can also be readily studied with ultracold particles.  The orthogonality catastrophe, where the many-body quantum ground state of system has vanishing overlap with its ground state in the presence of a localized interaction, figures prominently in the physics of the Kondo effect and the x-ray edge problem \cite{mahan}.  Here, we show that the orthogonality catastrophe also affects physisorption rates at low energies and modifies the probability that an ultracold atom will stick to a surface, a key quantity in many surface processes.  We predict a new energy-dependent scaling law for the sticking probability and show how this law differs quantitatively for the case of ultracold electrons adsorbed on the surface of porous silicon.  We then generalize our results for finite surface temperatures and find surprisingly the sticking probability changes sharply at a critical incident energy.  This singularity can be understood in terms of channel localization of the ultracold particles, a quantum phase transition that figures prominently in the dynamics of dissipative quantum systems \cite{leggett} at strong coupling.  

We start by considering a one-dimensional model that describes sticking for low energy particles at normal incidence.  The Hamiltonian is taken to be
\begin{equation}
H=H_p+H_b+H_c
\end{equation}
where
\begin{eqnarray}
H_p&=&{p^2\over 2m}+V_0(z),\\
H_b&=&\sum_q{\omega_q {a_q^\dagger} a_q},\\
H_c&=&-V_1(z)\sum_q \sigma\left(\omega_{q}\right) \ ({a_q+a_q^\dagger}) 
\end{eqnarray}

We might estimate the rate of sticking by one-phonon emission using Fermi's golden rule.  We first consider the transition matrix element $\langle b, q|H_c| k, 0\rangle$ of the dynamical particle-surface interaction, where $|k ,0\rangle$ denotes a state with the particle in the continuum state with incident energy $E={\hbar^2 k^2\over 2 m}$ and the surface in its ground state with no excitations; $|b, q\rangle$ denotes the particle in a bound state of $V$ and the surface has one excitation with wave number $q$.  

If we choose the normalization of the particle wave function such that it has unit amplitude far from the surface, for a potential $V_0(z)$ that decays faster than $z^{-2}$ as $z\to\infty$, it is well-known \cite{brenig,berlinsky,dpc92, hijmans,cole-review, dpc03} that the amplitude of the wave function near the surface scales as $k$.  This is a result of quantum reflection, a wave phenomenon where the incident particle is reflected from a surface without ever having reached a classical turning point.  Thus the amplitude of the wave function tends to vanish near the surface as the particle's incident energy tends to zero.  
We conclude that the transition matrix element for sticking  vanishes in proportion to $k$ at low energies. This result follows for all potentials that fall off faster than $z^{-2}$ for large $z$ and is universal in this sense.

Within lowest order perturbation theory, the sticking probability $s(E)$ of a particle with incident energy $E$ varies as the square of the transition matrix element and inversely with the incident particle flux. Hence, $s\propto {k^2/k}\sim \sqrt{E}$.  This low-energy threshold law for quantum sticking was implicit in work by Lennard-Jones \cite{lj3} in the pioneering years of quantum theory.  With improvements in the cooling and trapping techniques of ultracold atoms, this $\sqrt{E}$ threshold law was found to be consistent with experiment in the case of hydrogen sticking to the surface of superfluid helium  \cite{yu93}.  

It has been asserted \cite{george,cole-review,brenig} that the $\sqrt{E}$ law holds regardless of the form of $H_c$.  We will show that the $\sqrt{E}$ law only holds for a class of dynamical couplings $H_c$, determined by the low frequency behavior, in the same fashion as models of quantum dissipation are classified by their spectral functions.  For superohmic $H_c$, the $\sqrt{E}$ threshold law holds for neutral particles impinging on zero temperature surfaces; however, for ohmic couplings, we will show a different threshold law results.  In essence, $H_c$ contains a final-state interaction that can alter the ground state of the surface.  We have recently found that this final-state interaction is responsible for an orthogonality catastrophe for ohmic coupling \cite{dpc10} that subsequently alters the threshold law for quantum sticking.  

At sufficiently low energies, we can ignore inelastic scattering and approximate the particle state space by the initial state $|k \rangle$ and the final state $|b\rangle$.  In this truncated state space, the Hamiltonian becomes
\begin{eqnarray}
H&=&E c_k^\dagger c_k -E_b b^\dagger b+\sum_q{\omega_q {a_q^\dagger} a_q}
-(c_k^\dagger b+b^\dagger c_k)V_{kb}\sum_q \sigma\left(\omega_{q}\right) \ ({a_q+a_q^\dagger}) \nonumber\\
&&-c_k^\dagger c_k V_{kk}\sum_q \sigma\left(\omega_{q}\right)\ ({a_q+a_q^\dagger}) 
- b^\dagger b V_{bb}\sum_q \sigma\left(\omega_{q}\right) \ ({a_q+a_q^\dagger}) 
\label{ham}
\end{eqnarray}
where $V_{kb}=\langle k| V_1| b\rangle$ etc.  The effects of the $V_{kk}$ term in Eq.~\ref{ham} are of higher order in $k$ than the $V_{bb}$ term.  We consequently neglect the $V_{kk}$ term in what follows. 

 The $V_{bb}$ term is  the final-state interaction responsible for the orthogonality catastrophe for a certain class of frequency-dependent couplings.  The Hamiltonian for the surface excitations has a different form in the initial particle state compared to the final particle state.  Hence we need to include in the transition matrix element that the surface excitation in the final state is created from a different ground state from the initial ground state of the surface.  
 
The final state Hamiltonian of the surface, $H_{s,f}=\sum_q{\omega_q {a_q^\dagger} a_q}-V_{bb}\sum_q \sigma\left(\omega_{q}\right) \ ({a_q+a_q^\dagger})$, can be put in the form of that of the initial state by a displaced oscillator transformation.  Such a transformation reflects that the ground state of the surface in the presence of the bound particle is polarized relative to the surface in isolation.  To within an arbitrary phase factor, the overlap of the ground state of $H_{s,f}$ with that of the isolated surface is 
\begin{eqnarray}
S&\equiv&\langle 0_f| 0\rangle=e^{-F}\nonumber\\
&=&\exp{\left(-\frac{V^2_{bb}}{2}\sum_q {\sigma^2\left(\omega_{q}\right)\over\omega^2_q}\right)}
\label{fc}
\end{eqnarray}
In the continuum limit, this overlap vanishes when 
\begin{equation}
{\cal D}(\omega)\sigma^2(\omega)\sim\omega,\ \ \ \ \omega\to 0
\end{equation}
where ${\cal D}(\omega)$ is the density of surface excitations. In the language of models of quantum dissipation, this condition describes ohmic coupling \cite{leggett}. We have previously shown \cite{dpc10} this form of coupling applies to the dynamical interaction of the particle with phonons in an elastically isotropic surface; for example, in the case of Rayleigh phonons, $\sigma$ is independent of frequency \cite{flatte}, while ${\cal D}(\omega)\propto\omega$ for these two dimensional surface modes.  Hence, for sticking via the emission of Rayleigh phonons, the interaction is ohmic.  We also conclude that sticking via emission of bulk phonons (or ``mixed mode'' phonons) has an interaction that is ohmic, since ${\cal D}(\omega)\propto\omega^2$ and $\sigma\propto\omega^{-\frac{1}{2}}$.

In the golden rule estimate, the relevant transition matrix element should have an excitation created out of the final-state vacuum.  This reduces the transition matrix element by the Franck-Condon factor of Eq.~\ref{fc}, which in the ohmic case vanishes, signaling the breakdown of perturbation theory.  (We note that the dynamical coupling in the case of ultracold atomic hydrogen sticking to superfluid helium by ripplon emission is superohmic and consequently gives a non-vanishing Franck-Condon factor.  We expect the $\sqrt{E}$ law to hold in this case.)

In previous work \cite{dpc10}, we calculated  the sticking probability in the ohmic case using a non-perturbative variational scheme and found that the logarithmic divergence in the Franck-Condon factor is cut-off by a frequency scale $\omega_0$ that depends linearly on the incident particle energy $E$ at sufficiently low energies.  The low frequency cutoff $\omega_0$ might be thought of as coming from the finite time needed for the particle to make a transition to the bound state.  Excitations with frequencies below $\omega_0$ do not have adequate time to adjust to the presence of the bound particle and do not contribute to the Franck-Condon factor.    For weak $V_{bb}$,
\begin{equation}
F\approx\frac{\alpha}{2}\int_{c_1 E}^{\omega_c} {d\omega\over\omega}
\end{equation}
where $\alpha=\lim_{\omega\to 0}V^2_{bb}{{\cal D}(\omega)\sigma^2(\omega)/\omega}$, $c_1$ is a dimensionless constant, independent of $E$ and $\omega_c$ is the high-frequency cutoff of the bath.  

The truncation of the logarithmic divergence gives rise to a new behavior for the sticking probability at threshold for ohmic systems.
\begin{equation}
{\it s}(E)\propto E^{\frac{1}{2}}\cdot E^{\alpha}
\label{zerotneutral}
\end{equation}

We have considered a variety of experimental conditions to assess the likelihood that this new threshold law might be observed.  Unfortunately we have found that the shift in exponent $\alpha$ is typically much smaller than one.  Thus, this new threshold law might prove very challenging to verify experimentally.  However, in the case of charged particles sticking to surfaces, we are optimistic that the effects of the orthogonality catastrophe on the sticking probability will be accessible to experiment.  

The threshold law for charged particles differs from that of neutral particles.  Charged particles are influenced by a long-range attractive Coulomb interaction due to the particle's image charge, in contrast to the van der Waals interaction exerted on neutral particles.   The Coulomb potential decays sufficiently slowly that a low-energy charged particle does not experience quantum reflection \cite{dpc92}.  The amplitude of the wave function of the incident particle near the surface scales as $\sqrt{k}$ as $k\to 0$.  Hence, a na\"ive application of Fermi's golden rule would predict that the sticking probability $s(E)$ of a charged particle with incident energy $E$ behaves as $s\propto (\sqrt{k})^2/k\sim E^0$, a constant.  

For the case of an ohmic dynamical coupling, the orthogonality catastrophe modifies this na\"ive threshold law for charged particles.  The absence of quantum reflection for charged particles affects the energy-dependence of the low-frequency cutoff for the Franck-Condon factor, with $\omega_0$ scaling as $\sqrt{E}$ at low energies.  For small $\alpha$,
\begin{equation}
F\approx\frac{\alpha}{2}\int_{c_2\sqrt{E}}^{\omega_c} {d\omega\over\omega}
\end{equation}
where $c_2$ is a constant, independent of $E$.  
Thus, for charged particles, we obtain
\begin{equation}
{\it s}(E)\propto E^{\alpha/2}
\label{zerotcharged}
\end{equation}
In contrast to the na\"ive  threshold law where the sticking probability approaches a non-vanishing constant as $E\to 0$, the orthogonality catastrophe drives the sticking probability to zero.

It is a straightforward matter to extend this theory to surfaces at finite temperature, and there are several new features in the sticking probability that result.  The Franck-Condon factor is altered by thermally excited excitations in the bath
\begin{equation}
S=\exp{\left(-\frac{V^2_{bb}}{2}\sum_q {\sigma^2\left(\omega_{q}\right)\over\omega^2_q} \coth {\beta\omega_q\over 2}\right)}
\label{thermalfc}
\end{equation}

There is a critical incident energy $E_c$, dependent on the temperature $T$, below which the low-frequency cutoff $\omega_0$ sharply drops to zero.  As a result, the sticking probability is a victim of the orthogonality catastrophe and vanishes for $E< E_c$.  Consider the exponent $F$ of the Franck-Condon factor at finite temperature for vanishing cutoff $\omega_0$
 \begin{equation}
F\sim{\alpha} T \int_{0}{d\omega\over\omega^2}\to \infty
\end{equation}
Thus the Franck-Condon factor sharply drops to zero and the sticking probability vanishes for $E<E_c$, creating what might be termed a ``superreflective'' surface with perfect reflectivity below the critical incident energy.
This sharp change in the sticking probability has analogy with the behavior of the tunneling probability in the spin-boson model.  There, the tunneling probability is renormalized to zero beyond a temperature-dependent critical coupling to the bath, defining the localization phase boundary.

\begin{figure}
\includegraphics[width=12cm]{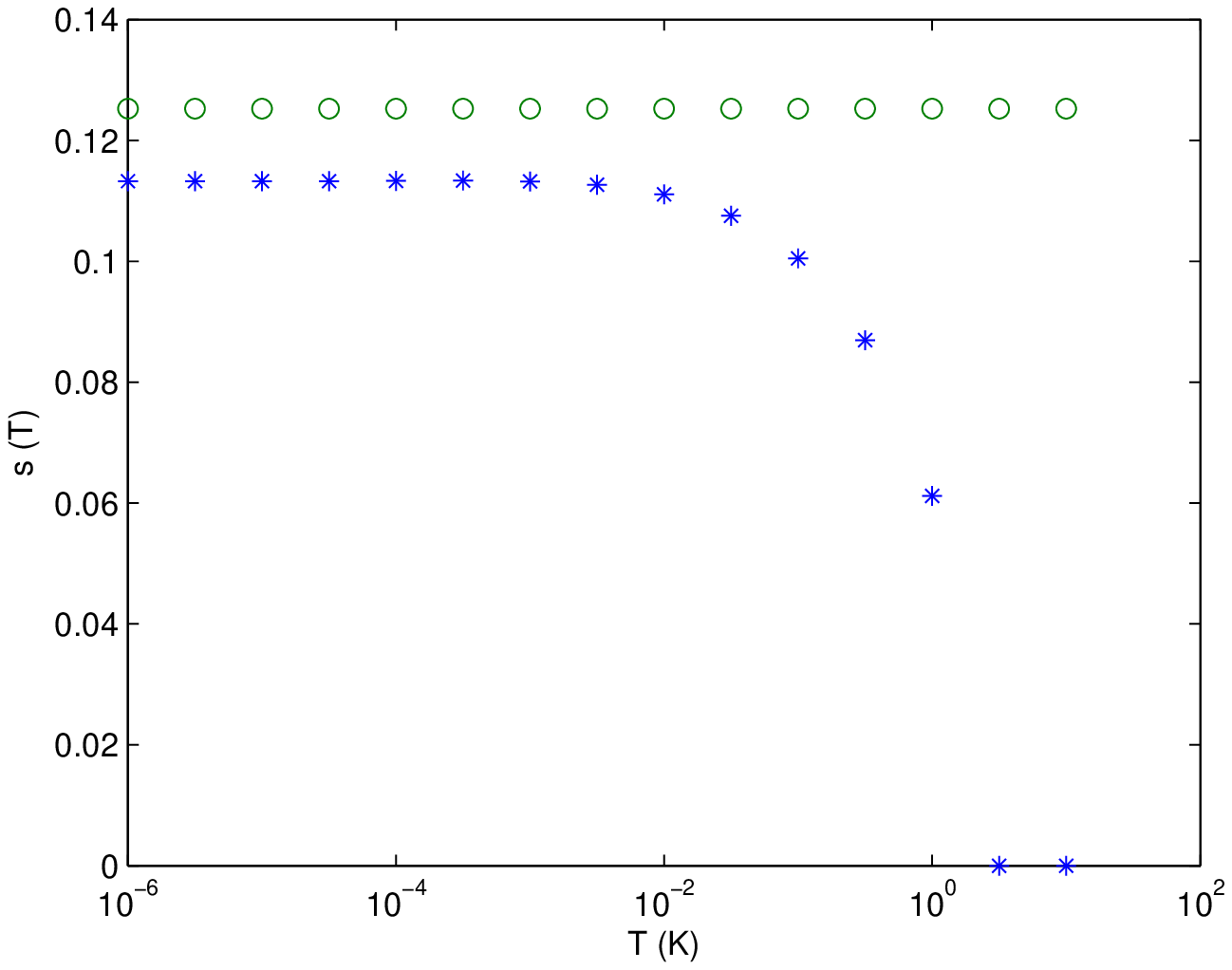}
\caption{\label{fig:svsT} (color online). The sticking probability of an electron ($E= 1$ mK) to the surface of porous silicon by the emission of a Rayleigh phonon as function of the surface temperature $T$.  The (green) circles result from using Fermi's golden rule with a Franck-Condon factor $S=1$.  The variational mean-field result is given by (blue) stars.  There is a dramatic downturn in the probability at a surface temperature of $T\approx 1.6$ K, corresponding to the vanishing of the Franck-Condon factor.}
\end{figure}

For the case of low surface temperature $T \ll \omega_0$, we recover the zero-temperature results of Eqs.~\ref{zerotneutral} and \ref{zerotcharged} to leading order in $T$; for the case of intermediate surface temperature where $T_c(E) > T \gg \omega_0$, we find that $F\propto \omega_0^{-1/2}$ for sufficiently low $\alpha$ and energy. However at finite temperature, our variational calculations show that $\omega_0\propto\sqrt{E}$ for neutral particles and $\sqrt[4]{E}$ for charged particles. 
Thus, we find for low-energy neutral particles with $E>E_c$ and intermediate surface temperatures
\begin{equation}
s(E) \propto \sqrt{E} \exp{\left(- \sqrt{E_0/E}\right)}
\end{equation}
while for low-energy charged particles,
\begin{equation}
s(E) \propto \exp{\left(-\sqrt[4]{E_0/{E}}\right)}
\end{equation}
where $E_0$ is an energy-independent constant and $T_c(E)$ is the critical temperature above which $\omega_0$ vanishes (see Fig.~\ref{fig:svsT}).

\begin{figure}
\includegraphics[width=12cm]{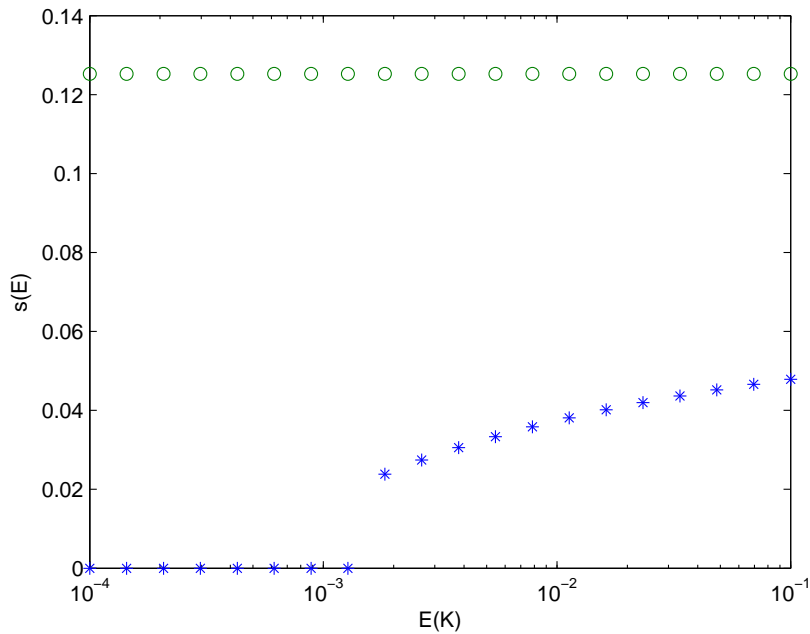}
\caption{\label{fig:svse} (color online). The sticking probability of an electron of energy $E$ to the surface of porous silicon by the emission of a Rayleigh phonon.  The surface temperature is taken to be $T= 2$ K.  The perturbative result using Fermi's golden rule with a Franck-Condon factor of 1 is given by (green) circles, while the variational mean-field result is given by (blue) stars.  The variational mean-field method gives a sharp transition at an incident energy $E\approx 1.6$ mK.  We take a porosity $P=92.9\%$, giving a dielectric constant $\kappa = 1.2$.  The shear modulus of $G=230$  MPa and Poisson's ratio $\sigma=0.03$ are calculated using Ref.~\cite {p-si} .}
\end{figure}

The numerical results from our variational mean-field theory for sticking are presented in Fig.~\ref{fig:svse} for the case of ultracold electrons sticking to porous silicon at finite temperature.  We choose highly porous silicon for two reasons:  to remain in the regime where sticking occurs predominantly through one-phonon processes, the binding energy must be small compared to $\omega_c$, the high frequency cutoff of the excitations; thus, we require a low dielectric constant.  Silicon with a porosity of $92.9\%$ has a dielectric constant of  only $\kappa = 1.2$.  Secondly, to maximize $\alpha$ in the case of coupling to Rayleigh phonons, we seek materials that have a low shear modulus and mass density.  We expect highly porous silicon to have a shear modulus of 230 MPa.

Our numerical calculations reveal a sharp transition in the sticking probability at a critical energy of $E_c\approx 1.6$ mK.  Electrons with energy below $E_c$ are predicted to be perfectly reflected by the surface.  Electrons with energy above $E_c$ stick to the surface with a probability reduced by roughly a factor of five compared to the na\"ive golden rule result.  

In summary, on the basis of a variational mean-field theory for the sticking of ultracold particles on a finite temperature surface, we predict new scaling laws of the sticking probability with incident energy at intermediate surface temperatures.  We also predict a dramatic downturn of the sticking probability, with the probability vanishing below a critical energy $E_c$.  This new feature in the sticking probability is a consequence of a bosonic orthogonality catastrophe, where the Franck-Condon factor, resulting from the surface polarization in the presence of the particle, vanishes for $E< E_c$.  We predict that this effect is experimentally accessible in the case of low energy electrons impinging on porous silicon.

\begin{acknowledgments}
We gratefully acknowledge support of this work by the National Science Foundation  under DMR-0814377. 
\end{acknowledgments}

\bibliography{qs}

\end{document}